\documentclass[9pt,twocolumn,twoside]{opticajnl}
\journal{opticajournal} 

\setboolean{shortarticle}{true}

\title{On-chip Kerr parametric oscillation with integrated heating for enhanced frequency tuning and control}

\author[1, *]{Jordan Stone}
\author[1]{Daron Westly}
\author[1,2]{Gregory Moille}
\author[1,2]{Kartik Srinivasan}

\affil[1]{Microsystems and Nanotechnology Division, Physical Measurement Laboratory, National Institute of Standards and Technology, Gaithersburg, MD 20899, USA}
\affil[2]{Joint Quantum Institute, NIST/University of Maryland, College Park, MD 20742, USA}

\affil[*]{jordan.stone@nist.gov}

\begin{abstract}
Nonlinear microresonators can convert light from chip-integrated sources into new wavelengths within the visible and near-infrared spectrum. For most applications, such as the interrogation of quantum systems with specific transition wavelengths, tuning the frequency of converted light is critical. Nonetheless, demonstrations of wavelength conversion have mostly overlooked this metric. Here, we apply efficient integrated heaters to tune the idler frequency produced by Kerr optical parametric oscillation in a silicon-nitride microring across a continuous $1.5$ terahertz range. Finally, we suppress idler frequency noise between $DC$ and $5$ kHz by several orders of magnitude using feedback to the heater drive.
\end{abstract}

\setboolean{displaycopyright}{false} 

\begin{document}

\maketitle

Nonlinear wavelength conversion, including harmonic and sum/difference frequency generation as well as optical parametric oscillation (OPO), in nanophotonic resonators is an energy-efficient way to generate coherent light at useful colors for a myriad of applications \cite{guo2016second, hao2017sum, Lu_Optica_2019, ledezma2023octave, surya2018efficient, lu2021ultralow}. In these devices, large resonator quality factors (Q) boost the effective nonlinearity and gain; however, the gain bandwidth is fundamentally constrained by the resonator linewidth. Hence, to tune the output frequency, one must control the resonator mode frequencies. Of course, frequency tunability is an indispensable aspect of laser systems, enabling spectral alignment with other optical systems (e.g., atomic transitions) for spectroscopy, frequency stabilization, and more. 
	
In particular, due to its intrinsic wavelength flexibility and efficiency, OPO based on four-wave mixing (FWM) in Kerr microresonators is being developed to address the visible-to-near-infrared spectrum \cite{lu2020chip, domeneguetti2021parametric, sayson2019octave}; but without sophisticated means to control the mode spectrum, frequency tuning has been limited to tens of gigahertz and further suffers from mode hopping \cite{sayson2019octave, tang2020widely, stone2022conversion}, where the signal and idler waves switch to adjacent longitudinal resonator modes instead of being continuously tuned. While mode hopping provides a convenient way to coarsely tune the OPO frequencies \cite{sun2024advancing, pidgayko2023voltage}, it must be suppressed for continuous tuning. Here, we demonstrate Kerr OPO with an unprecedented idler continuous frequency tuning range of $1.5$ THz. We leverage two recent advances in nonlinear microresonator technology: integrated heaters for efficient control of the mode spectrum \cite{moille2022integrated, nitiss2023tunable}, and wavenumber-selective OPO in photonic-crystal microresonators to prevent mode hopping \cite{stone2023wavelength}. We show that a targeted frequency deterministically maps to the microresonator temperature, and feedback to the heater drive enables frequency stabilization with in-loop noise reduced by more than six orders of magnitude. Our work explores a new regime of ultra-tunable nonlinear devices that will benefit both $\chi^{(2)}$ and $\chi^{(3)}$-type light sources.

\begin{figure}[t]
\centering
\includegraphics[width=250 pt]{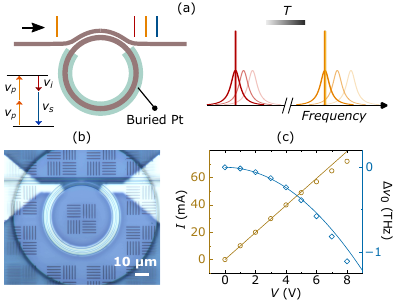}
\caption{(a) Illustration of Kerr microresonator optical parametric oscillation (OPO) and its temperature-based frequency control. (Left) Current passing through a buried platinum (Pt) strip generates heat flow to the microring, increasing its temperature, $T$. (Right) As a result, the microring mode spectrum (illustrated by Lorentzian-shaped curves) is shifted to lower frequencies (light-to-dark curves), along with any resonant lightwaves circulating the device. (b) Device micrograph. (c) Heater current, $I$ (gold circles), and OPO pump mode frequency shift, $\Delta \nu_{\rm{0}}$ (blue diamonds), versus the potential difference, $V$, between the buried heater contacts. Gold and blue curves are linear and quadratic fits to the data, respectively. }
\label{fig:1}
\end{figure}

In our experiments, we pump a silicon nitride (Si$_{3}$N$_{4}$, hereafter SiN) photonic-crystal microring near $795$ nm. The photonic crystal geometry has recently become a popular and powerful design choice for controlling dispersion and enabling novel nonlinear optical phenomena \cite{black2022optical, yu2021spontaneous, marty2021photonic,stone2023wavelength}. In our test device, the SiN layer thickness, ring outer radius, mean ring width, and ring width modulation amplitude are $450$ nm, $25$ $\mu$m, $820$ nm, and $10$ nm, respectively. Intraresonator FWM converts pump energy into signal and idler waves with widely-separated frequencies, and all three waves are resonant with fundamental transverse electric modes, so that temperature changes which shift the mode spectrum are conveyed to the OPO spectrum, as illustrated in Fig. \ref{fig:1}(a). Heat is transferred to the microring by driving current through a platinum (Pt) strip buried in the silicon dioxide (SiO$_2$) substrate $3$ $\mu$m below the SiN layer, as shown in Fig. \ref{fig:1}(a) and further described in Ref. \cite{moille2022integrated}. Figure \ref{fig:1}(b) shows a micrograph of the device used in our experiments. To characterize the heater performance, we measure its resistance and heating efficiency. Specifically, we measure the current flow, $I$, versus an applied potential difference, $V$, and we calculate the resistance $R=(100\pm1$) Ohms from a linear fit to these data (see Fig. \ref{fig:1}(c)). We also record the pump mode frequency shift, $\Delta \nu_{\rm{0}}$, where $\nu_{\rm{0}}\approx385$ THz, using laser transmission spectroscopy and an optical spectrum analyzer (OSA), and we observe a red-shift of more than 1~THz for the maximum applied voltage.

\begin{figure}[t]
\centering
\includegraphics[width=250 pt]{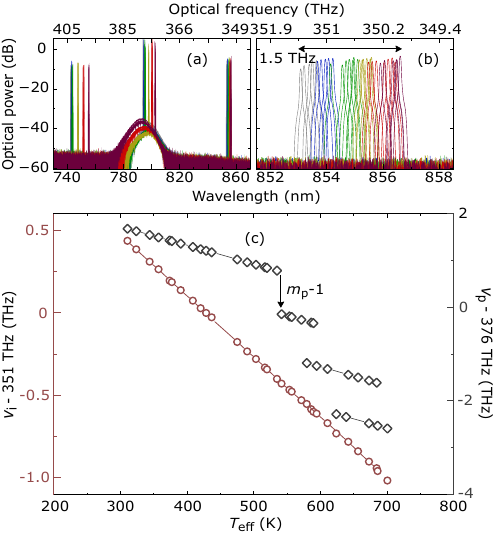}
\caption{(a) Compilation of OPO spectra from a single device. The six different colors (more evident in (b)) indicate six different $V$ setpoints: $0$, $4$, $5$, $6$, $7$, and $8$ V (left to right in the spectrum). At each setpoint, the pump laser frequency, $\nu_{\rm{p}}$, is scanned through a cavity resonance, and OPO spectra are periodically recorded. Here, $0$~dB is referenced to $1$~mW. (b) Spectra in (a) zoomed into the idler band. (c) Idler frequency, $\nu_{\rm{i}}$, and $\nu_{\rm{p}}$ (red circles and black diamonds, respectively) versus the estimated device temperature, $T_{\rm{eff}}$. Discontinuous jumps in $\nu_{\rm{p}}$ indicate the pump mode number, $m_{\rm{p}}$, is reduced by one to compensate the temperature-dependent dispersion.}
\label{fig:2}
\end{figure}

Our test device oscillates when pumped above its parametric threshold ($\approx30$ mW), generating an idler wave near $853$ nm with on-chip output power $\approx2$ mW. To examine frequency tuning, we periodically record the OPO spectrum while slowly scanning the pump laser frequency, $\nu_{\rm{p}}$, from shorter to longer wavelengths. We repeat this procedure for six different $V$ setpoints and present the optical spectra in Figs. \ref{fig:2}(a)-(b). From these spectra, we extract the idler frequencies, $\nu_{\rm{i}}$. Notably, for each $V$ setpoint, the range of $\nu_{\rm{i}}$ values measured during the $\nu_{\rm{p}}$ scan is more than $200$ GHz, which is much greater than the resonator linewidth ($\approx 1$ GHz). There are two reasons that we can access such a large range without changing $V$. First, the $\nu_{\rm{p}}$ scan induces large, dynamically-stable temperature increases which redshift the mode spectrum through the thermo-optic effect \cite{carmon2004dynamical}. (Hence, $\nu_{\rm{i}}$ is still being thermally tuned, but the temperature is increased via pump laser absorption rather than dissipated heater current). Second, the idler mode is fixed by the wavenumber-selective frequency matching scheme, so mode hopping does not occur during the $\nu_{\rm{p}}$ scan. Figure \ref{fig:2}(c) shows $\nu_{\rm{i}}$ and $\nu_{\rm{p}}$ versus the approximated device temperature, $T_{\rm{eff}}$, that we calculate as $T_{\rm{eff}}=T_{\rm{0}}+(\nu_{\rm{p}}-\nu_{\rm{0}}(T_{\rm{0}}))\times d\nu_{\rm{0}}/dT$, where $T_{\rm{0}}$ is the ambient temperature, $\nu_{\rm{0}}$ is the pump mode frequency at ambient temperature, and $d\nu_{\rm{0}}/dT\approx -4.5$ GHz/K is the measured pump mode thermal tuning coefficient. We achieve a continuous (i.e., free of spectral gaps) $\nu_{\rm{i}}$ tuning range of $1.5$ THz using at most $\approx500$ mW of electrical heating power (Fig.~\ref{fig:2}(b)). The relationship between $\nu_{\rm{i}}$ and $T_{\rm{eff}}$ is remarkably linear, indicating their deterministic connection that is enforced by large resonator $Q$. 

\begin{figure}[t]
\centering
\includegraphics[width=240 pt]{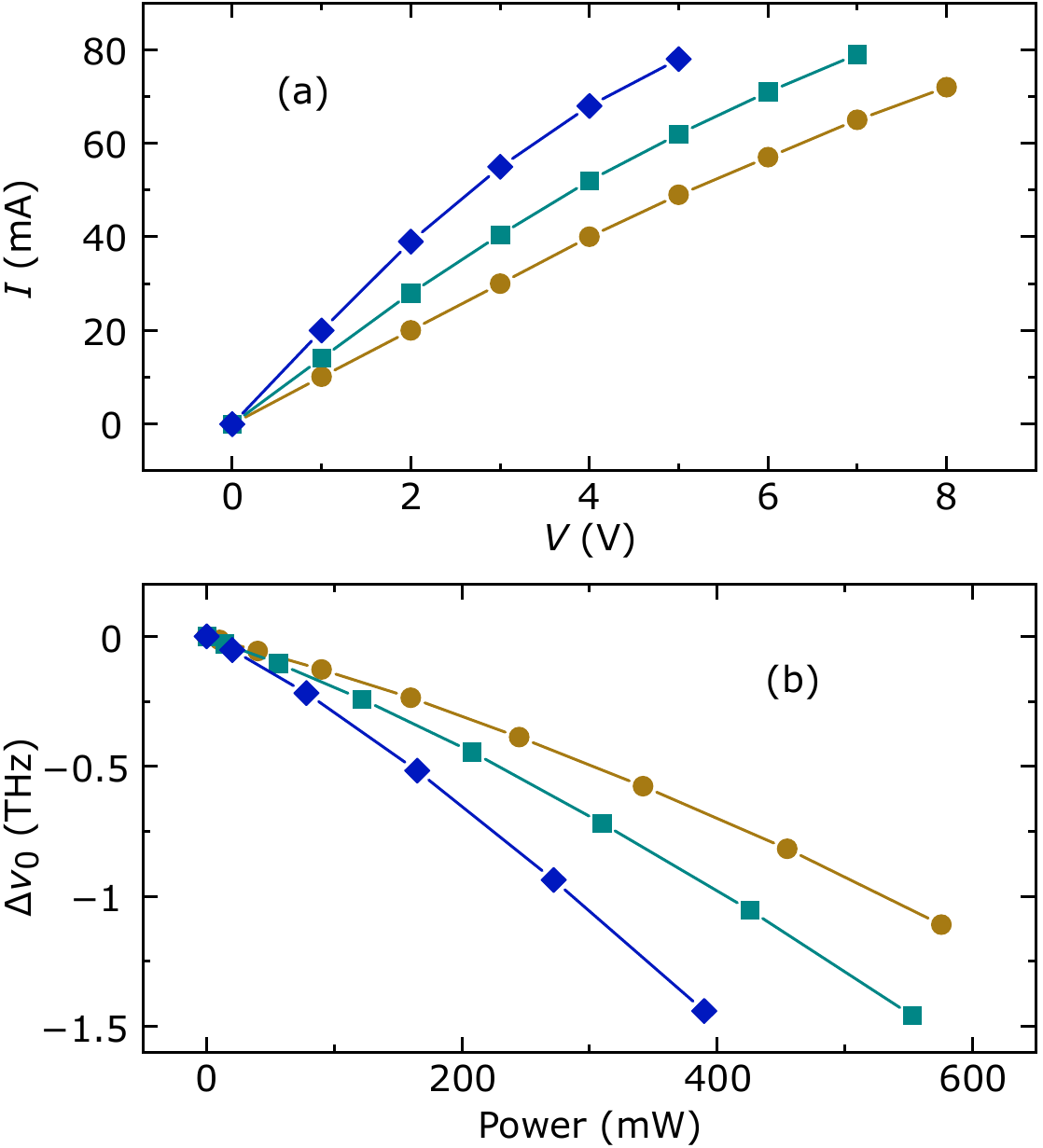}
\caption{(a) $I$ versus $V$ for three devices with different separations between the microresonator and heater leads. Device one (gold circles) is our OPO test device and is $\approx 1200$ $\mu$m from its electrical leads, while devices two (green squares) and three (blue diamonds) are $\approx 750$ $\mu$m and $\approx 200$ $\mu$m from their respective leads. The fitted low-$V$ resistance values for devices two and three are ($75 \pm 1$) Ohms and ($53 \pm 1$) Ohms, respectively. (b) $\Delta \nu_{\rm{0}}$ versus electrical heating power for the three devices in (a).}
\label{fig:3}
\end{figure}

In our test device, the Pt electrical leads and resonator are located in different sections of the chip; hence, there is a substantial electrical path length for which dissipated heat is not conducted to the resonator. While in future devices we anticipate increasing the cross-sectional area of the leads to lower their resistance, within the current process, shortening them will improve the heating efficiency. We test this hypothesis by characterizing two more devices and comparing them to the OPO test device (in these additional devices, the dispersion is not suitable to OPO; hence, we cannot directly compare their OPO tuning characteristics). In Fig. \ref{fig:3}(a), we present $I-V$ measurements for each device. We find that smaller separations between a microresonator and its electrical leads results in both decreased $R$ and a more nonlinear $I-V$ relationship. Next, we measure $\Delta \nu_{\rm{0}}$ versus the electrical heating power for each device and present the results in Fig. \ref{fig:3}(b). The device with the smallest resonator-lead separation (blue diamonds) is roughly $2\times$ more efficient than the OPO test device (gold circles).  

There are also other factors that determine the OPO tuning efficiency. For instance, the thermal tuning coefficient for a resonator mode with frequency $\nu_{\rm{m}}$ is $\frac{d\nu_{\rm{m}}}{dT} = \frac{\nu_{\rm{m}}}{n}\frac{dn}{dT}$, where $n$ is the refractive index and $dn/dT$ is the thermo-optic coefficient. Therefore, although we have chosen to examine $\nu_{\rm{i}}$ tuning due to the large output power and available analysis tools in the near-infrared, the OPO signal frequency, $\nu_{\rm{s}}$, is in general more tunable (such a device would implement wavenumber-selective frequency matching for the signal mode, instead of the idler mode as done here). For example, consider OPO for Na spectroscopy with $\nu_{\rm{s}}\approx 589$ nm. For a microresonator geometry similar to our test device, $\nu_{\rm{s}}$ could be tuned more than $2$ THz for the same temperature shifts implied in Fig. \ref{fig:2}.  

Notably, as $T_{\rm{eff}}$ is increased, the microring dispersion changes (primarily due to chromatic dispersion of the thermo-optic coefficient) and eventually prevents oscillation when pumping a given mode. However, we compensate by decrementing the pump longitudinal mode number, $m_{\rm{p}}$, by one, as marked in Fig. \ref{fig:2}(c). In fact, the total $\nu_{\rm{i}}$ tuning range is limited by the $\nu_{\rm{p}}$ tuning range (due to decreasing gain of the pump laser amplifier at longer wavelengths); with more $\nu_{\rm{p}}$ tunability, we expect $\nu_{\rm{i}}$ could be tuned further. Importantly, the idler mode is fixed by the wavenumber-selective frequency matching scheme \cite{stone2023wavelength}, so switching $m_{\rm{p}}$ does not impact the $\nu_{\rm{i}}$ tunability (on the other hand, the signal mode number is decreased by two, as seen in Fig. \ref{fig:2}(a)). Still, future pump laser integration may confine the range of $\nu_{\rm{p}}$ values; hence, the $\nu_{\rm{i}}$ tuning range obtained without $m_{\rm{p}}$ adjustments is an important metric. In our experiments, this value is $\approx 820$ GHz, which is approaching the $\approx900$ GHz free spectral range of our device, and the corresponding electrical power is $200$ mW. We expect that thermally-induced dispersion shifts can be mitigated through advanced dispersion engineering techniques, e.g., by controlling the spatial profiles of microresonator modes to balance thermo-optic dispersion.  

\begin{figure}[t]
\centering
\includegraphics[width=250 pt]{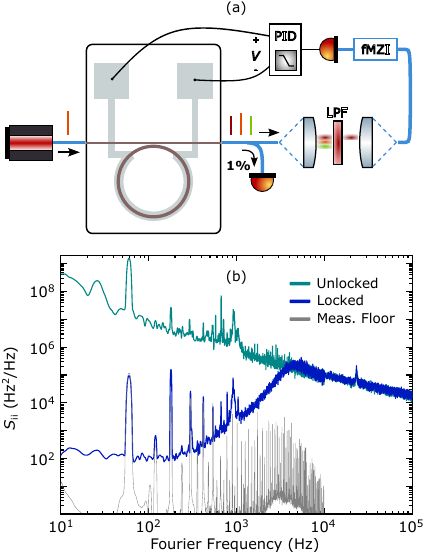}
\caption{(a) Experimental schematic for controlling the idler frequency through feedback to the heater drive voltage, $V$. The output idler wave is coupled from the chip access waveguide into fiber and isolated from the pump and signal waves using a free-space filtering setup. Prior to filtering, one percent of the output light is tapped and detected to monitor the transmitted power. Then, idler frequency fluctuations are transduced to power fluctuations using a fiber-based Mach-Zehnder interferometer (fMZI). Detection of the fMZI output yields an error signal for processing and feedback. LPF: Long-pass filter; PID: Proportional-integral-derivative control. (b) Idler frequency noise spectra with (blue) and without (green) feedback to $V$. The gray curve indicates the measurement floor.}
\label{fig:4}
\end{figure}

Finally, we explore $\nu_{\rm{i}}$ stabilization through feedback to $V$. We filter the device output to isolate the idler wave and use a fiber Mach-Zehnder interferometer to transduce $\nu_{\rm{i}}$ fluctuations into optical power fluctuations that we detect with a transimpedance-amplified photodiode whose output serves as an error signal. In Fig. \ref{fig:4}(a), we provide the experimental schematic. The error signal is processed with proportional-integral-derivative control, and we measure its power spectral density using an electronic spectrum analyzer. In Fig. \ref{fig:4}(b), we plot the in-loop (i.e., using our error signal monitor) power spectral density of $\nu_{\rm{i}}$ fluctuations, $\mathcal{S}_{\rm{ii}}$. Without feedback, the dominant contribution to $\mathcal{S}_{\rm{ii}}$ at offset frequencies less than $1$ kHz is pump laser frequency noise. Above this frequency, we assess that intrinsic thermorefractive noise plays a large role \cite{huang2019thermorefractive}. With feedback, we reduce $\mathcal{S}_{\rm{ii}}$ by more than six orders of magnitude at low offset frequencies, and we observe noise suppression at offset frequencies up to $5$ kHz. Notably, $V$ feedback does not influence $\nu_{\rm{p}}$, so $\nu_{\rm{i}}$ stabilization is enabled for any pump laser technology. 

In conclusion, we have used integrated heating to realize efficient and agile frequency control of nonlinear nanophotonics. We experimentally focused on Kerr OPO, where integrated heating dramatically increased the range of continuous frequency tuning to $1.5$ THz, and feedback enabled substantial broadband noise reduction. We emphasize that our heating/tuning scheme is broadly applicable to other resonantly-enhanced nonlinear processes, such as harmonic generation. In that context, self-injection-locked microresonators have recently been developed to support full system integration \cite{clementi2023chip, ling2023self, shen2020integrated} and would benefit from integrated control of the mode spectrum. Finally, we note that our OPO platform exhibits sufficient output power, phase noise, and frequency tunability to replace bulkier light sources holding back the deployment of quantum technologies. Future work will fulfill this mission by developing pump laser- and microelectronics-integrated systems.

\begin{backmatter}
\bmsection{Funding} We acknowledge funding support from the DARPA LUMOS and NIST-on-a-chip programs. 

\bmsection{Acknowledgments} We are grateful to Dave Long and Oscar Ou for their feedback during manuscript preparation, and we thank Yi Sun for assistance with the micrograph. 


\smallskip

\bmsection{Disclosures} DW, GM, and KS have filed a provisional patent application on the buried heater technology used in this work.

\bmsection{Data availability} Data underlying the results presented in this paper may be obtained from the authors upon reasonable request.

\bigskip


\end{backmatter}




\vspace{-0.5cm}
\bibliography{references}


\bibliographyfullrefs{references}




\end{document}